\documentclass[review, 12pt]{elsarticle}
\usepackage{lineno,hyperref,amsmath,amsthm}
\usepackage[noend]{algpseudocode}
\usepackage{algorithmicx,algorithm}
\usepackage{booktabs}
\usepackage{color}
\usepackage{graphics}
\usepackage{subfig}
\usepackage{amsfonts,amssymb}
\usepackage{bm}

\modulolinenumbers[1]

\journal{Journal} %

\begin{document}

\begin{frontmatter}

\title{A sample-based stochastic finite element method for structural reliability analysis}

\author[a,b]{Zhibao Zheng\corref{CorrespondingAuthor}}
\cortext[CorrespondingAuthor]{Corresponding author}
\ead{zhibaozheng@hit.edu.cn}

\author[a,b]{Hongzhe Dai}

\author[a,b]{Yuyin Wang}

\author[a,b]{Wei Wang}

\address[a]{Key Lab of Structures Dynamic Behavior and Control, Harbin Institute of Technology, Ministry of Education, Harbin 150090, China}

\address[b]{School of Civil Engineering, Harbin Institute of Technology, Harbin 150090, China}

\begin{abstract} 
This paper presents a new methodology for structural reliability analysis via stochastic finite element method (SFEM). A novel sample-based SFEM is firstly used to compute structural stochastic responses of all spatial points at the same time, which decouples the stochastic response into a combination of a series of deterministic responses with random variable coefficients, and solves corresponding stochastic finite element equation through an iterative algorithm. Based on the stochastic response obtained by the SFEM, the limit state function described by the stochastic response and the multidimensional integral encountered in reliability analysis can be computed without any difficulties, and failure probabilities of all spatial points are calculated once time. The proposed method can be applied to high-dimensional stochastic problems, and one of the most challenging issues encountered in high-dimensional reliability analysis, known as Curse of Dimensionality, can be circumvented without expensive computational costs. Three practical examples, including large-scale and high-dimensional reliability analysis, are given to demonstrate the accuracy and efficiency of the proposed method in comparison to the Monte Carlo simulation.
\end{abstract}


\end{frontmatter}


\section{Introduction}
As a powerful tool to quantify the uncertainty in practical problems, reliability analysis nowadays has become an indispensable cornerstone for solving complex problems in many fields \cite{melchers2018structural, thoft2012structural}, such as structural design and optimization, decision management, etc. Despite the progress in existing methods of modeling and analysis, the estimation  of the failure probability in reliability analysis is challenging to achieve \cite{koutsourelakis2004reliability, au1999new}. On one hand, the multidimensional integral encountered in reliability analysis for computing the failure probability often lies in high-dimensional stochastic spaces (hundreds to more), which is prohibited because of expensive computational costs. On another hand, the failure region, that is, the region of unacceptable system performance, is usually complicated and irregular, which leads that the failure surface is rarely known explicitly and only can be evaluated by numerical solutions.

Many methods are proposed in order to evaluate multidimensional integrals arising in reliability analysis. The most straightforward method is the Monte Carlo simulation (MCS), which is based on the law of large numbers and almost converges to the exact value when the number of samples is large enough \cite{papadrakakis2002reliability}. The MCS doesn't depend on the dimension of stochastic spaces, thus it doesn't encounter the Curse of Dimensionality. However, the computational cost for estimating a small failure probability is expensive, which is prohibited in practical complex problems. As a very robust technique, it is usually used to check the effectiveness of other methods. Some variations have been proposed to improve the MCS, such as importance sampling, subset simulation, etc \cite{au1999new, dubourg2011reliability, au2007application}. A typically kind of non-sampling methods for reliability analysis are First/Second Order Reliability Method (FORM/SORM) \cite{der1987second, sudret2000stochastic}. These method are based on first/second order series expansion approximation of the failure surface at the so-called design point, then the resulting approximate integral is calculated by asymptotic method. Thses methods generally have good accuracies and efficiencies for low-dimensional and weakly nonlinear problems, however considerable errors arise in high-dimensional stochastic spaces and nonlinear failure surfaces \cite{rackwitz2001reliability}. Several mehtods have been proposed to improve the performance of these methods \cite{schueller2004critical}. In order to decrease the computational cost for reliability analysis, surrogate model methods are receiving particular attention and continuously gaining in significance, which are based on a functional surrogate representation as an approximation of the limit state function. The surrogate model is constructed in an explicit expression via a set of observed points, then the failure probability can be estimated with cheap computational costs. The construction of the surrogate model is crucial, and available surrogate models include response surface method \cite{gavin2008high, li2011stochastic}, Kriging method \cite{Zhang2015Efficient}, support vector machine \cite{Bourinet2016Rare}, high dimensional model representation \cite{chowdhury2009hybrid}, polynomial chaos expansion \cite{li2010evaluation, marelli2018active, cheng2020structural}, etc.

In most practical cases, the limit state function in reliability analysis builds a relationship between stochastic spaces of input parameters and the failure probability via the stochastic response of the system \cite{li2010evaluation, Zuev2015Subset, Zhang2013Structural}, thus the determination of the stochastic response of the system is crucial. For decades, the stochastic finite element method (SFEM), especially the spectral stochastic finite element method and its extensions \cite{ghanem2003stochastic, Xiu2002The, Matthies2005Galerkin, Nouy2009Recent, stefanou2009stochastic}, has received particular attentions. As an extension of the classical deterministic finite element method to the stochastic framework, SFEM has been proven efficient both numerically and analytically on numerous problems in engineering and science \cite{xiu2010numerical}. In this method, the unknown stochastic response is projected onto a stochastic space spanned by (generalized) polynomial chaos basis, and stochastic Galerkin method is then adopted to transform the original finite element equation into a deterministic finite element equation, whose size can be up to orders of magnitude larger than that of the corresponding stochastic problems \cite{ghanem2003stochastic, Xiu2002The}. Extrme computational costs arise as the number of stochastic dimensions and the number of polynomial chaos expansion terms increase, thus the high resolution solution of SFEM is still challenging due to the increased memory and computational resources required, especially for high-dimensional and large-scale stochastic problems \cite{Nouy2009Recent, stefanou2009stochastic}.

In this paper, in order to overcome the difficulties encountered in existing SFEM, we adopt a novel sample-based stochastic finite element method \cite{zheng2020anovelsfem} to compute stochastic responses of target systems. In this method, the unknown stochastic response is expanded into a combination of a series of deterministic responses with random variable coefficients described by samples. More importantly, it can be applied to high-dimensional and large-scale stochastic problems with a high accuracy and efficiency. Based on the obtained stochastic response, the limit state function and the multidimensional integrals in reliability analysis can be computed without any difficulties, and failure probabilities of all spatial points are calculated once time without expensive computational costs.

The paper is organized as follows: Section \ref{Sec2} presents a novel sample-based stochastic finite element method for determining structural stochastic responses. Reliability analysis based on stochastic responses is described in Section \ref{Sec3}. Following this, the algorithm implementation of the proposed method is elaborated in Section \ref{Sec4}. Three practical problems are used to demonstrate good performances of the proposed method in Section \ref{Sec5}. Some conclusions and prospects are discussed in Section \ref{Sec6}.

\section{Stochastic responses determination using SFEM}\label{Sec2}
As an extension of deterministic finite element method (FEM), SFEM has become a common tool for computing structural stochastic responses \cite{ghanem2003stochastic, xiu2010numerical}. Modeling random system parameters and environmental sources by use of random fields \cite{phoon2002simulation, zheng2017simulation}, it becomes available to integrate discretization methods for structural responses and random fields to arrive at a system of stochastic finite element equations as
\begin{equation}\label{SFE}
  K\left( \theta  \right)u\left( \theta  \right) = F\left( \theta  \right)
\end{equation}
where $K\left( \theta  \right)$ is the stochastic global stiffness matrix representing properties of the physical model under investigation, $u\left( \theta  \right)$ is the unknown stochastic response and $F\left( \theta  \right)$ is the load vector associated with the source terms.

As one of the most important problems of SFEM, it's a great challenge to compute the high-precision solution of Eq.\eqref{SFE}. Spectral stochastic finite element method (SSFEM) is a popular method in the past few decades, which represents the stochastic response through polynomial chaos expansion (PCE) and transform Eq.\eqref{SFE} into a deterministic finite element equation by stochastic Galerkin projection \cite{Nouy2009Recent, xiu2010numerical}. The size of the deterministic finite element equation is much larger than that of the original problem, thus expensive computational costs limit SSFEM to low-dimensional stochastic problems. In order to overcome these difficulties, a novel sample-based SFEM is developed to solve Eq.\eqref{SFE} in \cite{zheng2020anovelsfem}, which represents the unknown stochastic response $u\left( \theta  \right)$ as
\begin{equation}\label{uk}
  u\left( \theta  \right) = \sum\limits_{i = 1}^k {{\lambda _i}\left( \theta  \right){d_i}}
\end{equation}
where $\left\{ {{\lambda _i}(\theta )} \right\}_{i = 1}^k$ and $\left\{ {{d_i}} \right\}_{i = 1}^k$ are unknown random variables and unknown deterministic vectors, respectivey. Solution $u\left( \theta  \right)$ is approximated after $k$ terms truncated, and the more terms $k$ retains, the more accurate approximation can be obtained. In order to compute the couple $\left\{ {{\lambda _k}\left( \theta  \right),{d_k}} \right\}$, supposing that the $k-1$ terms $\left\{ {{\lambda _i}\left( \theta  \right),{d_i}} \right\}_{i = 1}^{k - 1}$ have been obtained and substituting Eq.\ref{uk} into Eq.\eqref{SFE} yields,
\begin{equation}\label{LDk}
  K\left( \theta  \right)\left[ {\sum\limits_{i = 1}^{k - 1} {{\lambda _i}\left( \theta  \right){d_i}}  + {\lambda _k}\left( \theta  \right){d_k}} \right] = F\left( \theta  \right)
\end{equation}

It's not easy to determine ${\lambda _k}\left( \theta  \right)$ and $d_k$ at the same time. In order to avoid this difficulty, ${\lambda _k}\left( \theta  \right)$ and $d_k$ are computed one after another. For determined random variable ${\lambda _k}\left( \theta  \right)$ (or given as an initial value), $d_k$ can be computed by using stochastic Galerkin method, which corresponds to
\begin{equation}\label{Dk}
  E\left\{ {{\lambda _k}\left( \theta  \right)K\left( \theta  \right)\left[ {\sum\limits_{i = 1}^{k-1} {{\lambda _i}\left( \theta  \right){d_i}}  + {\lambda _k}\left( \theta  \right){d_k}} \right]} \right\} = E\left\{ {{\lambda _k}\left( \theta  \right)F\left( \theta  \right)} \right\}
\end{equation}
where $E\{  \cdot \}$ is the expectation operator. Once $d_k$ has been determined in Eq.\eqref{Dk}, the random variable ${\lambda _k}\left( \theta  \right)$ can be subsequently computed via multiplying $d_k$ on both sides of Eq.\eqref{LDk}. It yields
\begin{equation}\label{Lk}
  d_k^TK\left( \theta  \right)\left[ {\sum\limits_{i = 1}^{k-1} {{\lambda _i}\left( \theta  \right){d_i}}  + {\lambda _k}\left( \theta  \right){d_k}} \right] = d_k^TF\left( \theta  \right)
\end{equation}
The couple $\left\{ {{\lambda _k}\left( \theta  \right),{d_k}} \right\}$ can be computed by repeating Eq.\eqref{Dk} and Eq.\eqref{Lk} until it converges to the required accuracy. For the practical implementation, $d_k$ is unitized as $d_k^T{d_k} = 1$, and the convergence error of the couple $\left\{ {{\lambda _k}\left( \theta  \right),{d_k}} \right\}$ is defined as,
\begin{equation}\label{Loc_err}
  {\varepsilon _{local}} = \frac{{E\left\{ {{{\left( {{\lambda _{k,j}}\left( \theta  \right){d_{k,j}}} \right)}^2} - {{\left( {{\lambda _{k,j - 1}}\left( \theta  \right){d_{k,j - 1}}} \right)}^2}} \right\}}}{{E\left\{ {{{\left( {{\lambda _{k,j}}\left( \theta  \right){d_{k,j}}} \right)}^2}} \right\}}} = 1 - \frac{{E\left\{ {\lambda _{k,j - 1}^2\left( \theta  \right)} \right\}}}{{E\left\{ {\lambda _{k,j}^2\left( \theta  \right)} \right\}}}
\end{equation}
which measures the difference between ${\lambda _{k,j}}\left( \theta  \right)$ and ${\lambda _{k,j - 1}}\left( \theta  \right)$ and the calculation is stopped when ${\lambda _{k,j}}\left( \theta  \right)$ is almost the same as ${\lambda _{k,j - 1}}\left( \theta  \right)$. Further, the stop criterion of the number $k$ that are retained of the stochastic solution $u\left( \theta  \right)$ is defined as
\begin{equation}\label{Glo_err}
  {\varepsilon _{global}} = \frac{{E\left\{ {u_k^2\left( \theta  \right) - u_{k - 1}^2\left( \theta  \right)} \right\}}}{{E\left\{ {u_k^2\left( \theta  \right)} \right\}}} = 1 - \frac{{\sum\limits_{i,j = 1}^{k - 1} {E\left\{ {{\lambda _i}\left( \theta  \right){\lambda _j}\left( \theta  \right)} \right\}d_i^T{d_j}} }}{{\sum\limits_{i,j = 1}^k {E\left\{ {{\lambda _i}\left( \theta  \right){\lambda _j}\left( \theta  \right)} \right\}d_i^T{d_j}} }}
\end{equation}

In a practical way, the stochastic global stiffness matrix $K\left( \theta  \right)$ and stochastic global load vector $F\left( \theta  \right)$ in stochastic finite element equation Eq.\eqref{SFE} are obtained by assembling stochastic element stiffness matrices and stochastic element load vector as
\begin{equation}\label{KFe}
  K\left( \theta  \right) = \sum\limits_{l = 0}^M {{\xi _l}\left( \theta  \right){K_l}},~F\left( \theta  \right) = \sum\limits_{m = 0}^Q {{\eta _m}\left( \theta  \right){F_m}}
\end{equation}
Based on Eq.\eqref{KFe}, Eq.\eqref{Dk} can be simplified and written as
\begin{equation}\label{Dk2}
  {\tilde K_{kk}}{d_k} = \sum\limits_{m = 0}^Q {{h_{mk}}{F_m}}  - \sum\limits_{i = 1}^{k - 1} {{{\tilde K}_{ik}}{d_i}}
\end{equation}
where deterministic matrices ${\tilde K_{ij}}$ are given by
\begin{equation}\label{Kij}
  {\tilde K_{ij}} = \sum\limits_{l = 0}^M {{c_{lij}}{K_l}}
\end{equation}
and coefficients $c_{ijk}$ and $h_{ij}$ are computed by
\begin{equation}\label{Cij}
  {c_{ijk}} = E\left\{ {{\xi _i}\left( \theta  \right){\lambda _j}\left( \theta  \right){\lambda _k}\left( \theta  \right)} \right\},~{h_{ij}} = E\left\{ {{\eta _i}\left( \theta  \right){\lambda _j}\left( \theta  \right)} \right\}
\end{equation}

The size of ${\tilde K_{ij}}$ in Eq.\eqref{Kij} is the same as the original stochastic finite element equation Eq.\eqref{SFE}, which can be solved by existing deterministic FEM techniques \cite{hughes2012finite, reddy2014introduction}, thus it is readily applied to large-scale stochastic problems. Similarly, Eq.\eqref{Lk} can be simplified and written as
\begin{equation}\label{ab}
  {a_k}\left( \theta  \right){\lambda _k}\left( \theta  \right) = {b_k}\left( \theta  \right)
\end{equation}
where random variables ${a_k}\left( \theta  \right)$ and ${b_k}\left( \theta  \right)$ are given by
\begin{equation}\label{ABk}
  {a_k}\left( \theta  \right) = \sum\limits_{l = 0}^M {{g_{klk}}{\xi _l}\left( \theta  \right)} ,~ {b_k}\left( \theta  \right) = \sum\limits_{m = 0}^Q {{f_{km}}{\eta _m}\left( \theta  \right)}  - \sum\limits_{i = 1}^{k - 1} {\sum\limits_{l = 0}^M {{g_{kli}}{\xi _l}\left( \theta  \right){\lambda _i}\left( \theta  \right)} }
\end{equation}
and coefficients $g_{ijk}$ and $f_{ij}$ are computed by
\begin{equation}\label{Gij}
  {g_{ijk}} = d_i^T{K_j}{d_k} ,~ {f_{ij}} = d_i^T{F_j}
\end{equation}

The common methods solving Eq.\eqref{ab} by representing the random variable $\lambda_k(\theta)$ in terms of a set of polynomial chaos have expensive computational costs \cite{Nouy2009Recent, xiu2010numerical}. In order to avoid this difficulty, a sample-based method is adopted to determine $\lambda_k(\theta)$. For sample realizations $\{ {\theta ^{(n)}}\} _{n = 1}^N$ of all considered random parameters $\theta$, sample matrices of random variables ${a_k}\left( \theta  \right)$ and ${b_k}\left( \theta  \right)$ are written as
\begin{equation}\label{SamAB}
  {{\bm{\tilde a}}_k}\left( \theta  \right) = {\bm{\tilde \xi }}\left( \theta  \right){{\bf{g}}_{k, \cdot ,k}} ,~ {{\bm{\tilde b}}_k}\left( \theta  \right) = {\bm{\tilde \eta }}\left( \theta  \right){{\bf{f}}_k} - \left( {{\bm{\tilde \xi }}\left( \theta  \right){{\bf{g}}_{k, \cdot ,1:k - 1}} \odot {{{\bm{\tilde \lambda }}}^{\left( {k - 1} \right)}}\left( \theta  \right)} \right){\left[ {\bf{1}} \right]_{\left( {k - 1} \right) \times 1}}
\end{equation}
where ${{\bm{\tilde a}}_k}\left( \theta  \right),~{{\bm{\tilde b}}_k}\left( \theta  \right) \in {\mathbb{R}^{N \times 1}}$, $\odot$ reprents element-by-element multiplication of ${\bm{\tilde \xi }}\left( \theta  \right){{\bf{g}}_{k, \cdot ,1:k - 1}}$ and ${{{\bm{\tilde \lambda }}}^{\left( {k - 1} \right)}}\left( \theta  \right)$, and the sample matrices of random variables $\left\{ {{\xi _i}\left( \theta  \right)} \right\}_{i = 0}^M$, $\left\{ {{\eta _i}\left( \theta  \right)} \right\}_{i = 0}^Q$, $\left\{ {{\lambda _i}\left( \theta  \right)} \right\}_{i = 1}^{k - 1}$ are given by
\begin{equation}\label{SamK1}
  {\bm{\tilde \xi }}\left( \theta  \right) = \left[ {\begin{array}{*{20}{c}}
  	1&{{\xi _1}\left( {{\theta ^{\left( 1 \right)}}} \right)}& \cdots &{{\xi _M}\left( {{\theta ^{\left( 1 \right)}}} \right)}\\
  	\vdots & \vdots & \ddots & \vdots \\
  	1&{{\xi _1}\left( {{\theta ^{\left( N \right)}}} \right)}& \cdots &{{\xi _M}\left( {{\theta ^{\left( N \right)}}} \right)}
  	\end{array}} \right] \in {\mathbb{R}^{N \times \left( {M + 1} \right)}}
\end{equation}
\begin{equation}\label{SamK2}
  {\bm{\tilde \eta }}\left( \theta  \right) = \left[ {\begin{array}{*{20}{c}}
  	1&{{\eta _1}\left( {{\theta ^{\left( 1 \right)}}} \right)}& \cdots &{{\eta _Q}\left( {{\theta ^{\left( 1 \right)}}} \right)}\\
  	\vdots & \vdots & \ddots & \vdots \\
  	1&{{\eta _1}\left( {{\theta ^{\left( N \right)}}} \right)}& \cdots &{{\eta _Q}\left( {{\theta ^{\left( N \right)}}} \right)}
  	\end{array}} \right] \in {\mathbb{R}^{N \times \left( {Q + 1} \right)}}
\end{equation}
\begin{equation}\label{SamK3}
  {{\bm{\tilde \lambda }}^{\left( {k - 1} \right)}}\left( \theta  \right) = \left[ {\begin{array}{*{20}{c}}
  	{{\lambda _1}\left( {{\theta ^{\left( 1 \right)}}} \right)}& \cdots &{{\lambda _{k - 1}}\left( {{\theta ^{\left( 1 \right)}}} \right)}\\
  	\vdots & \ddots & \vdots \\
  	{{\lambda _1}\left( {{\theta ^{\left( N \right)}}} \right)}& \cdots &{{\lambda _{k - 1}}\left( {{\theta ^{\left( N \right)}}} \right)}
  	\end{array}} \right] \in {\mathbb{R}^{N \times \left( {k - 1} \right)}}
\end{equation}
and the coefficient matrices are obtained by
\begin{equation}\label{MG}
  {{\bf{g}}_k} = \left[ {{g_{kij}}} \right] \in {\mathbb{R}^{\left( {M + 1} \right) \times k}},~{{\bf{f}}_k} = \left[ {{f_{km}}} \right] \in {\mathbb{R}^{\left( {Q + 1} \right) \times 1}}
\end{equation}
where ${\bf{g}}_{k, \cdot ,k} \in {\mathbb{R}^{\left( {M + 1} \right) \times 1}}$ represents the $k$-th column of the matrix ${{\bf{g}}_k}$ and ${\bf{g}}_{k, \cdot ,1:k - 1} \in {\mathbb{R}^{\left( {M + 1} \right) \times {k-1}}}$ represents the 1-st column to $\left( {k - 1} \right)$-th column of the matrix ${{\bf{g}}_k}$. By use of the sample realizations ${{\tilde a}_k}\left( \theta  \right)$ and ${{\tilde b}_k}\left( \theta  \right)$, sample realizations ${{\bm{\tilde \lambda }}_k}\left( \theta  \right)$ of the random variable ${{\lambda _k}\left( \theta  \right)}$ can be obtained by
\begin{equation}\label{SamL}
  {{\bm{\tilde \lambda }}_k}\left( \theta  \right) = \frac{{{{{\bm{\tilde a}}}_k}\left( \theta  \right)}}{{{{{\bm{\tilde b}}}_k}\left( \theta  \right)}}
\end{equation}

Statistics methods are readily introduced to obtain probability characteristics of the random variable $\lambda_k(\theta)$ from samples ${{\bm{\tilde \lambda }}_k}\left( \theta  \right)$. The computational cost for solving Eq.\eqref{SamL} mainly comes from  computing the sample vectors ${{\bm{\tilde a}}_k}\left( \theta  \right)$ and ${{\bm{\tilde b}}_k}\left( \theta  \right)$ in Eq.\eqref{SamAB}. It is very low even for high-dimensional stochastic problems, that is, Eq.\eqref{SamAB} is insensitive to the dimensions of ${\bm{\tilde \xi }}\left( \theta  \right)$ and ${\bm{\tilde \eta }}\left( \theta  \right)$, which avoid the Curse of Dimensionality to great extent. Hence, the proposed method is particularly appropriate for high-dimensional stochastic problems in practice.

\section{Reliability analysis}\label{Sec3}
Reliability analysis is typically described by a scalar limit state function $g\left( \theta  \right)$ and corresponding failure probability $P_F$, which requires the evaluation of the following multidimensional integral \cite{koutsourelakis2004reliability, au1999new}
\begin{equation}\label{PfO}
  {P_F} = \int_{g\left( \theta  \right) \le 0} {f\left( \theta  \right)d\theta }
\end{equation}
where $g\left( \theta  \right) \le 0$ is the failure domain, $f\left( \theta  \right)$ represents the joint probability density function (PDF) of random variables associated with system parameters and environmental sources. The integral Eq.\eqref{PfO} for determining the failure probability is usually difficult to evaluate since the limit state surface or failure surface $g\left( \theta  \right) = 0$ has a very complicated geometry and $f\left( \theta  \right)$ is defined in high-dimensional stochastic spaces. In most cases the form of the limit state function $g\left( \theta  \right)$ is not known explicitly and numerical methods are employed for the evaluation of Eq.\eqref{PfO}. Existing reliability analysis methods generally evaluate the failure probability of a single point. For general purpose, the spatial limit state function $g\left( {{\bf{x}},\theta } \right)$ is considered as a vector function of spatial positions $\bf{x}$. Similarly, the spatial failure probability function ${P_F}\left( {\bf{x}} \right)$ is defined as
\begin{equation}\label{Pfx}
  {P_F}\left( {\bf{x}} \right) = \int_{g\left( {{\bf{x}},\theta } \right) \le 0} {f\left( {{\bf{x}},\theta } \right)d\theta }
\end{equation}

Due to the introduction of spatial positions $\bf{x}$, the failure probability function ${P_F}\left( {\bf{x}} \right)$ in Eq.\eqref{Pfx} is more difficult to compute than that in Eq.\eqref{PfO}. In fact, $g\left( {{\bf{x}},\theta } \right)$ is typically represents a complicated relation between the inputs and the failure modes via the solution of a potential highly complex stochastic system. Here, we compute the solution of the complex stochastic system by use of SFEM mentioned in Section \ref{Sec2}. Considering the stochastic response $u\left( \theta  \right) = \sum\limits_{i = 1}^k {{\lambda _i}\left( \theta  \right){d_i}}$ and substituting it into $g\left( {{\bf{x}},\theta } \right)$ yield
\begin{equation}\label{Gx}
  g\left( {{\bf{x}},\theta } \right) = g\left( {\sum\limits_{i = 1}^k {{\lambda _i}\left( \theta  \right){d_i}} } \right)
\end{equation}
where the random parameters of the system are integrated in the random variables $\left\{ {{\lambda _i}\left( \theta  \right)} \right\}_{i = 1}^k$ and the spatial parameter $\bf{x}$ is discretized and embedded in the deterministic vectors $\left\{ {{d_i}} \right\}_{i = 1}^k$. Thus, the failure probability function ${P_F}\left( {\bf{x}} \right)$ in Eq.\eqref{Pfx} can be rewritten as
\begin{equation}\label{PfC}
  {P_F}\left( {\bf{x}} \right) = \Pr \left[ {g\left( {\sum\limits_{i = 1}^k {{\lambda _i}\left( \theta  \right){d_i}} } \right) \le 0} \right]
\end{equation}

The most straightforward and efficient way to compute Eq.\eqref{PfC} is Monte Carlo simulation and its variations, where random samples are generated according to the distribution of $\theta$. Numbers of the points land in the failure domain are counted to estimate the failure probability. Similar to the process of Monte Carlo simulation, we utilize a sample-based method to estimate Eq.\eqref{PfC}. Random samples $\left\{ {{\lambda _i}\left( {{\theta ^{\left( n \right)}}} \right)} \right\}_{n = 1}^N$ ($N$ is the number of samples) in Eq.\eqref{PfC} have been generated by use of Eq.\eqref{SamL}, thus the failure probability function ${P_F}\left( {\bf{x}} \right)$ can be evaluated in the following form
\begin{equation}\label{PfS}
  {P_F}\left( {\bf{x}} \right) = \frac{1}{N}\sum\limits_{n = 1}^N {I\left[ {g\left( {\sum\limits_{i = 1}^k {{\lambda _i}\left( {{\theta ^{\left( n \right)}}} \right){d_i}} } \right)} \right]}
\end{equation}
where $I\left(  \cdot  \right)$ is the indicator function satisfying
\begin{equation}\label{In}
  I\left( s \right) = \left\{ \begin{array}{l}
  1,~s \le 0\\
  0,~s > 0
  \end{array} \right.
\end{equation}

The proposed method in Eq.\eqref{PfS} combines the high accuracy of sampling methods and the high efficiency of non-sampling methods. On one hand, as a sample-based method, it has a comparable accuracy with Monte Carlo method, and the accuracy increases as the number of samples increases. On the other, it doesn't require a full-scale simulation of the underlying system for each sample point and only depends on the solution obtained by SFEM, thus the computing efficiency is greatly improved. In addition, Eq.\eqref{PfS} can compute the failure probability ${P_F}\left( x_i \right)$ for each spatial point $x_i$ once time, which provides a simple but effective way to identify multiple failure modes of complex structures. Hence, the proposed method provides an efficient and unified framework for reliability analysis, and is particularly appropriate for high-dimensional and complex stochastic problems in practice.

\section{Algorithm implementation}\label{Sec4}

\begin{algorithm}[h]
	\caption{}
	\label{alg1}
	\begin{algorithmic}[1]
		\State Initialize samples of the random variables $\left\{ {{\xi _l}\left( {{\theta ^{\left( n \right)}}} \right)} \right\}_{n = 1}^N,~l = 1, \cdots ,M$ and $\left\{ {{\eta _m}\left( {{\theta ^{\left( n \right)}}} \right)} \right\}_{n = 1}^N,~m = 1, \cdots ,Q$;
		\label{Step01}    
		\While {${\varepsilon _{global}} > {\varepsilon _1}$}
		\label{Step02}    
		\State Initialize samples of the random variable $\left\{ {{\lambda _{k,0}}\left( {{\theta ^{\left( n \right)}}} \right)} \right\}_{n = 1}^N$;
		\label{Step03}    
		\Repeat
		\label{Step04}    
		\State Compute the response component ${d_{k,j}}$ by solving Eq.\eqref{Dk2};
		\label{Step05}    
		\State Compute the random variable ${\lambda _{k,j}}\left( \theta  \right)$ via Eq.\eqref{SamL};
		\label{Step06}    
		\Until ${\varepsilon _{local}} < {\varepsilon _2}$
		\label{Step07}    
		\State ${u_k}\left( \theta  \right) = \sum\limits_{i = 1}^{k - 1} {{\lambda _i}\left( \theta  \right){d_i} + {\lambda _k}\left( \theta  \right){d_k}},\; k \ge 2$;
		\label{Step08}    
		\EndWhile
		\State ${\bf{end}}$ ${\bf{while}}$
		\label{Step09}    
		\State Compute the spatial limit state function $g\left( {{\bf{x}},\theta } \right)$ via Eq.\eqref{Gx};
		\label{Step10}    
		\State Compute the spatial failure probability function ${P_F}\left( {\bf{x}} \right)$ via Eq.\eqref{PfS};
		\label{Step11}    
	\end{algorithmic}
\end{algorithm}

The resulting procedures for solving the stochastic finite element equation Eq.\eqref{SFE} and computing the the failure probability function ${P_F}\left( {\bf{x}} \right)$ via Eq.\eqref{Pfx} are summarized in Algorithm \ref{alg1}, which includes two parts in turn. The first part is from step \ref{Step02} to step \ref{Step09}, which is to compute the stochastic response $u\left( \theta  \right)$ and includes a double-loop iteration procedure. The inner loop, which is from step \ref{Step04} to step \ref{Step07}, is used to determine the couple of $\left\{ {{\lambda _k}(\theta ),{d_k}} \right\}$, while the outer loop, which is from step \ref{Step02} to step \ref{Step09}, corresponds to recursively building the set of couples such that the approximate solution $u_k(\theta)$ satisfies Eq.\eqref{SFE}. In step \ref{Step02} and step \ref{Step07}, iteration errors $\varepsilon _{global}$ and $\varepsilon _{local}$ are calculated via Eq.\eqref{Glo_err} and Eq.\eqref{Loc_err}, and corresponding convergence errors $\varepsilon _1$ and $\varepsilon _2$ are required precisions. The second part includes step \ref{Step10} and step \ref{Step11}, where the limit state function $g\left( {{\bf{x}},\theta } \right)$ is generated in step \ref{Step10} based on the stochastic response $u\left( \theta  \right)$ obtained in step \ref{Step08} and the failure probability function ${P_F}\left( {\bf{x}} \right)$ is computed in step \ref{Step11}.

\section{Numerical examples} \label{Sec5}
\begin{figure}[!b]
	\begin{center}
		\includegraphics[width=0.45\textwidth]{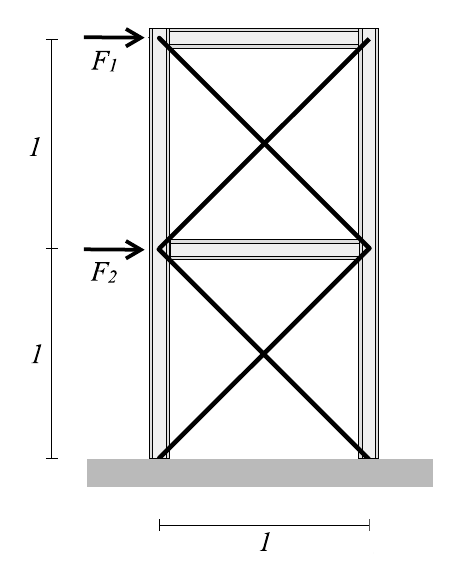}
		\caption{Model of the two-layer frame.}
		\label{e1_model}
	\end{center}
\end{figure}
In this section, we present three examples, including the reliability analysis of a beam-bar frame, the reliability analysis of a roof truss defined in 100-dimensional stochastic spaces and the global reliability analysis of a plate, to illustrate the accuracy and efficiency of the proposed method in comparison to $1 \times {10^{6}}$ times Monte Carlo simulations. For all considered examples, $1 \times {10^{6}}$ initial samples for each random variable $\left\{ {{\xi _l}\left( {{\theta ^{\left( n \right)}}} \right)} \right\}_{n = 1}^{1 \times {{10}^6}}$, $\left\{ {{\eta _m}\left( {{\theta ^{\left( n \right)}}} \right)} \right\}_{n = 1}^{1 \times {{10}^6}}$ and $\left\{ {{\lambda _{k,0}}\left( {{\theta ^{\left( n \right)}}} \right)} \right\}_{n = 1}^{1 \times {{10}^6}}$ are generated, and the convergence errors in step \ref{Step02} and step \ref{Step07} of Algorithm \ref{alg1} are set as ${\varepsilon _1} = 1 \times {10^{ - 5}}$  and ${\varepsilon _2} = 1 \times {10^{ - 3}}$, respectively.

\subsection{Reliability analysis of a beam-bar frame} \label{Example1}
\begin{table}[!b]
	\centering
	\caption{Probability distributions of random variables in the Example \ref{Example1}.}
	\label{e1_rv}
	\begin{tabular}{llllr}
		\toprule
		variable & description & distribution & mean & variance \\
		\midrule
		$E_{beam}$ & Young’s modulus of beam & normal & 210~MPa & 0.2\\
		$A_{beam}$ & cross-sectional area of beam & lognormal & 100~mm$^2$ & 0.2\\
		$I _{beam}$ & moment of inertia of beam & lognormal & 800~mm$^4$ & 0.2\\
		$E_{bar}$ & Young’s modulus of bar & normal & 210~MPa & 0.2\\
		$A_{bar}$ & cross-sectional area of bar & lognormal  & 100~mm$^2$ & 0.2\\
		$F_1$ & load 1 & normal & 10~kN & 0.2\\
		$F_2$ & load 2 & normal & 10~kN  & 0.2\\
		\bottomrule
	\end{tabular}
\end{table}
A two-layer frame is shown in Fig.\ref{e1_model}, which consists of horizontal and vertical beams, and is stabilized with diagonal bars. Probability distributions of independent random variables associated with material properties, geometry properties and loads are listed in Table \ref{e1_rv}. In this example, we consider the failure probability of a single point and the limit state function $g\left( \theta  \right)$ is defined by the maximum joint displacement of the frame as
\begin{equation}\label{e1_gx}
  g\left( \theta  \right) = \mathop {\max }\limits_i \sqrt {u_{{x_i}}^2 + u_{{y_i}}^2}  - c \cdot {u_{mean}}
\end{equation}
where ${u_{mean}} = {\rm{mean}}\left( {\mathop {\max }\limits_i \sqrt {u_{{x_i}}^2\left( \theta  \right) + u_{{y_i}}^2\left( \theta  \right)} } \right)$ is the mean value of the maximum joint displacement, and the scalar $c$ is related to different failure probabilities, that is, the failure probability decreases as the scalar $c$ increases. In this paper, the maximum joint displacement of the frame can be identified automatically by the proposed method instead of selecting manually, since the proposed method can calculate the stochastic response of all nodes once time.

\begin{figure}[!b]
	\begin{center}
		\includegraphics[width=0.7\textwidth]{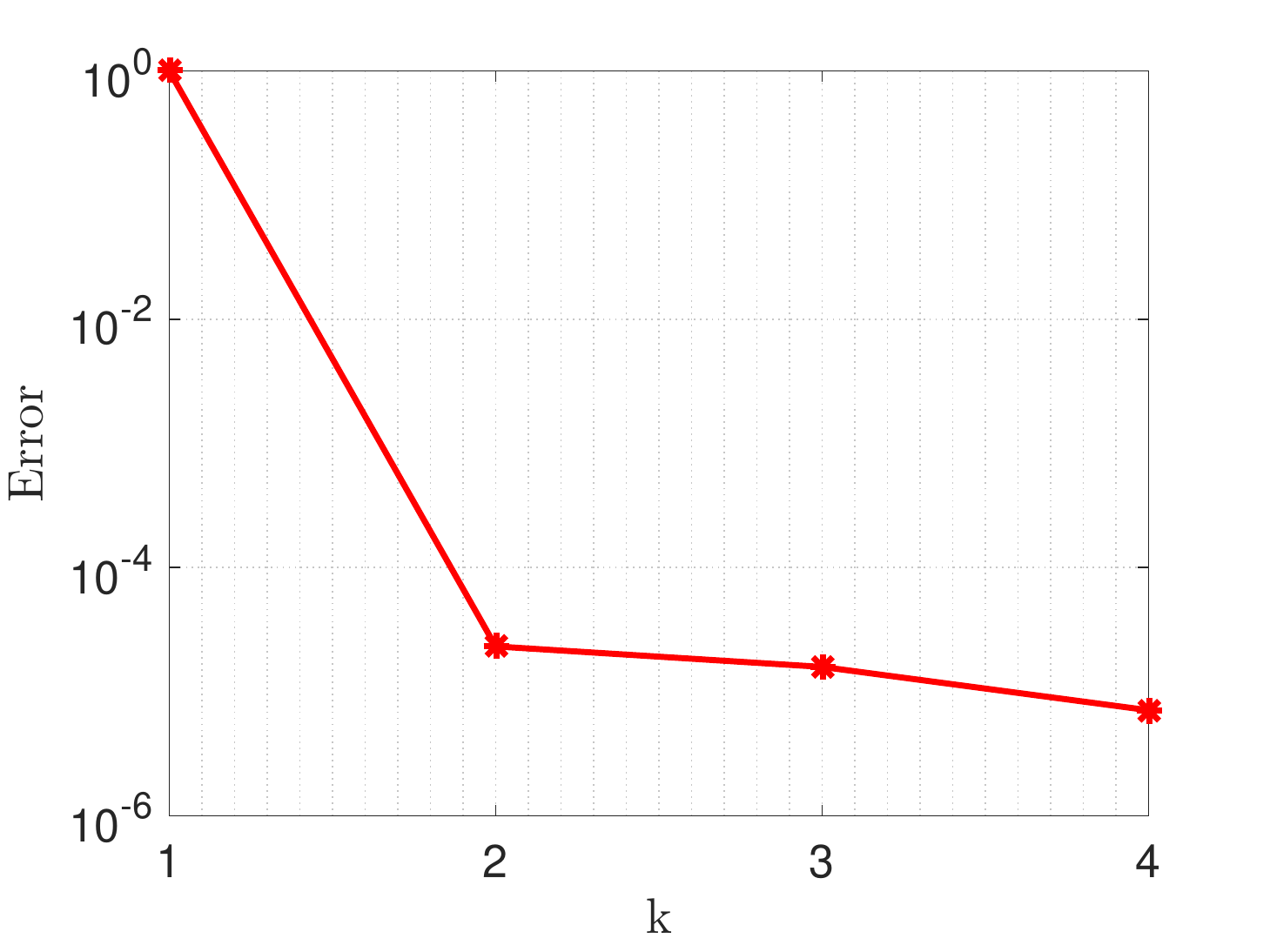}
		\caption{Iteration errors of different retained items.}
		\label{e1_err}
	\end{center}
\end{figure}
\begin{figure}[!t]
	\begin{center}
	\includegraphics[width=1.0\textwidth]{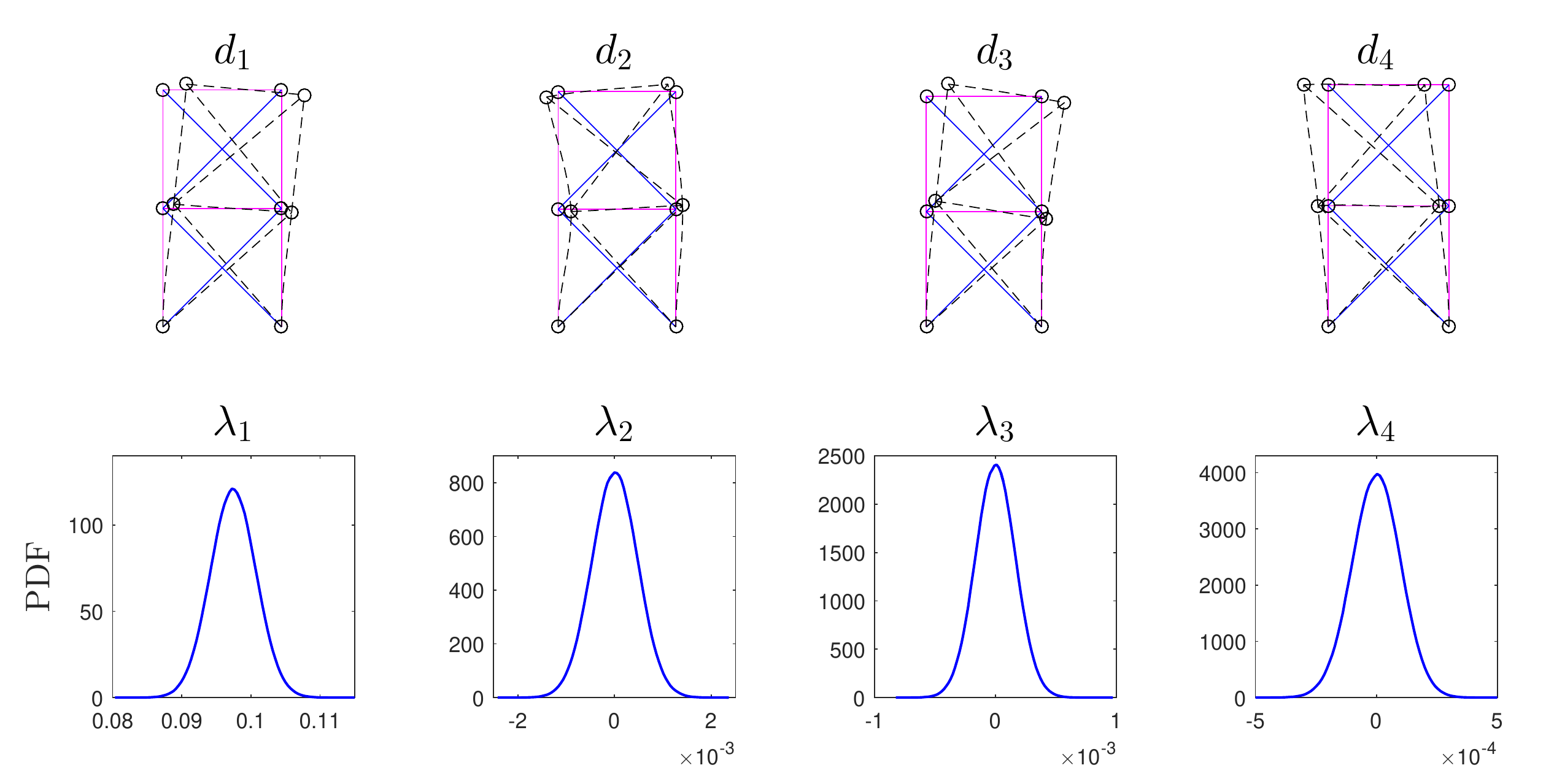}
	\caption{Solutions of the couples $\left\{ {{\lambda _i}\left( \theta  \right),{d_i}} \right\}_{i = 1}^4$.}
	\label{e1_DL}
	\end{center}
\end{figure}
\begin{figure}[!t]
	\begin{center}
		\includegraphics[width=0.7\textwidth]{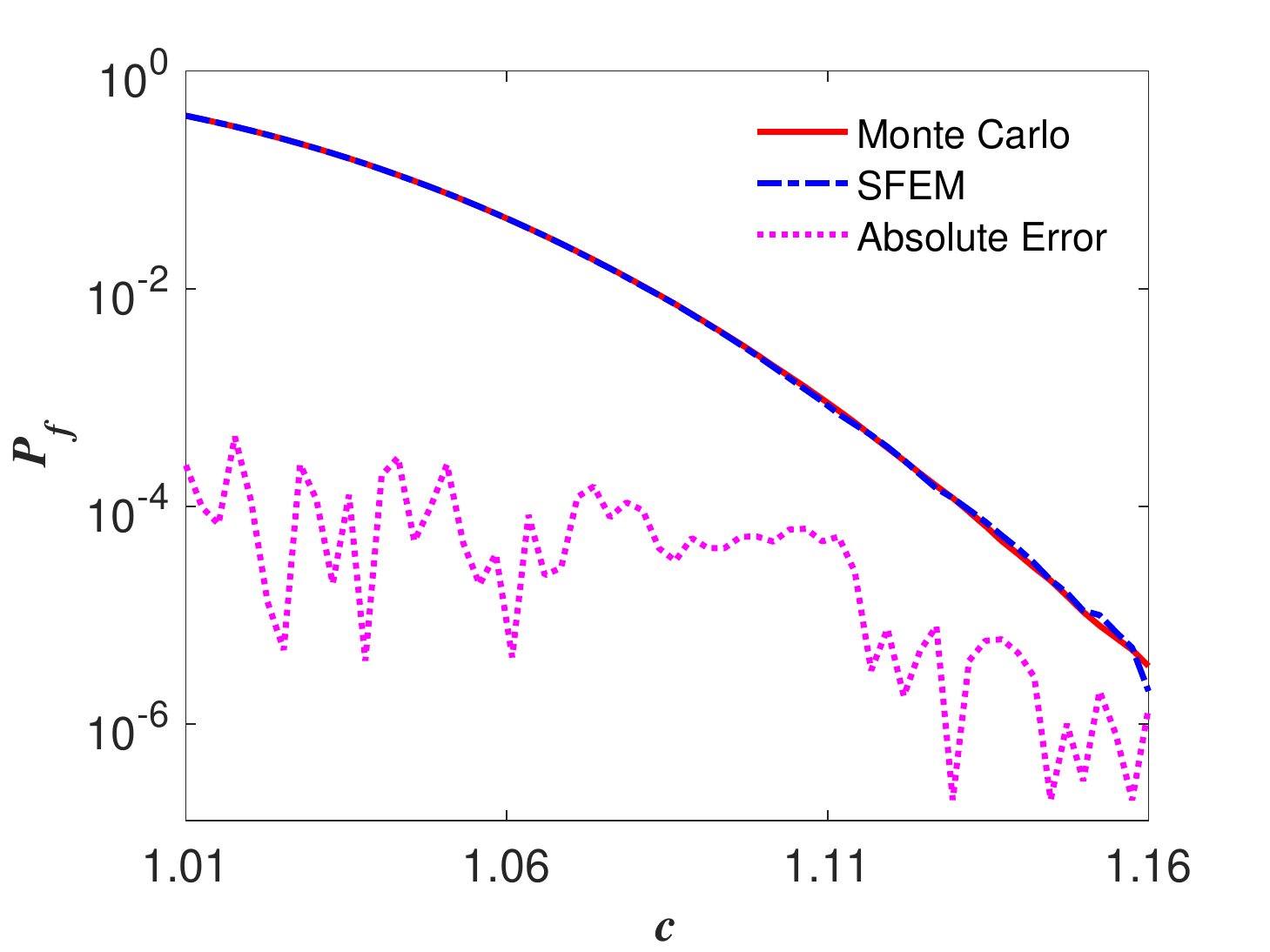}
		\caption{Failure probabilities of different scalar $c$.}
		\label{e1_Pf}
	\end{center}
\end{figure}

In order to compute the failure probability $P_f$, we firstly compute the stochastic response of the frame by using the first part of Algorithm \ref{alg1}. It is seen from Fig.\ref{e1_err} that only 4 iterations can achieve the required precision ${\varepsilon _1} = 1 \times {10^{ - 5}}$, which demonstrates the fast convergence rate of the proposed method. Correspondingly, the number of couples $\left\{ {{\lambda _k}\left( \theta  \right),{d_k}} \right\}$ that constitute the stochastic response is adopted as $k=4$, as shown in Fig.\ref{e1_DL}. With the increasing of the number of couples, the ranges of corresponding random variables are more closely approaching to zero, which indicates that the contribution of the higher order random variables to the approximate solution decays dramatically. 

Based on the stochastic response obtained by SFEM, failure probabilities of different scalar $c$ are shown in Fig.\ref{e1_Pf}, here the scalar $c$ is set from 1.01 to 1.16. The failure probability $P_f$ computed from the proposed method ranges from $10^0$ to $10^{-6}$, which is fairly close to that obtained from the Monte Carlo simulation even for a very small failure probability. The absolute error between the proposed method and MC simulation demonstrate the accuracy and efficiency of the proposed method.

\subsection{Reliability analysis of a roof truss} \label{Example2}
\begin{figure}[!b]
	\begin{center}
		\includegraphics[width=0.7\textwidth]{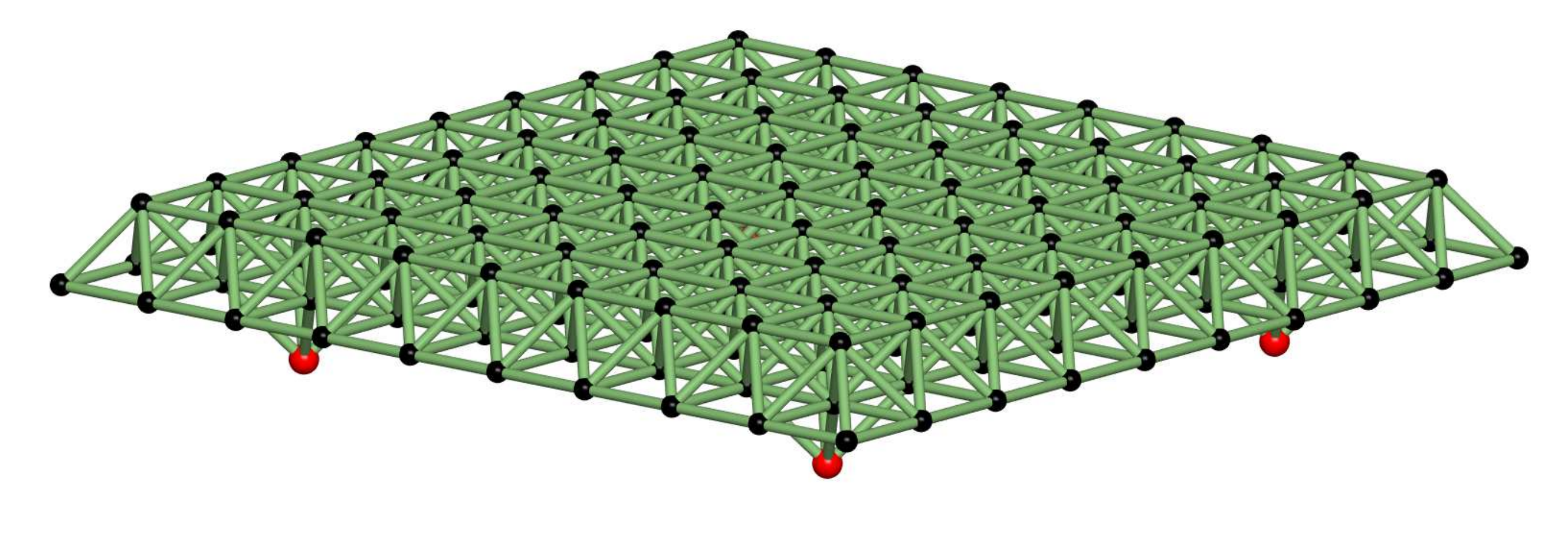}
		\includegraphics[width=0.7\textwidth]{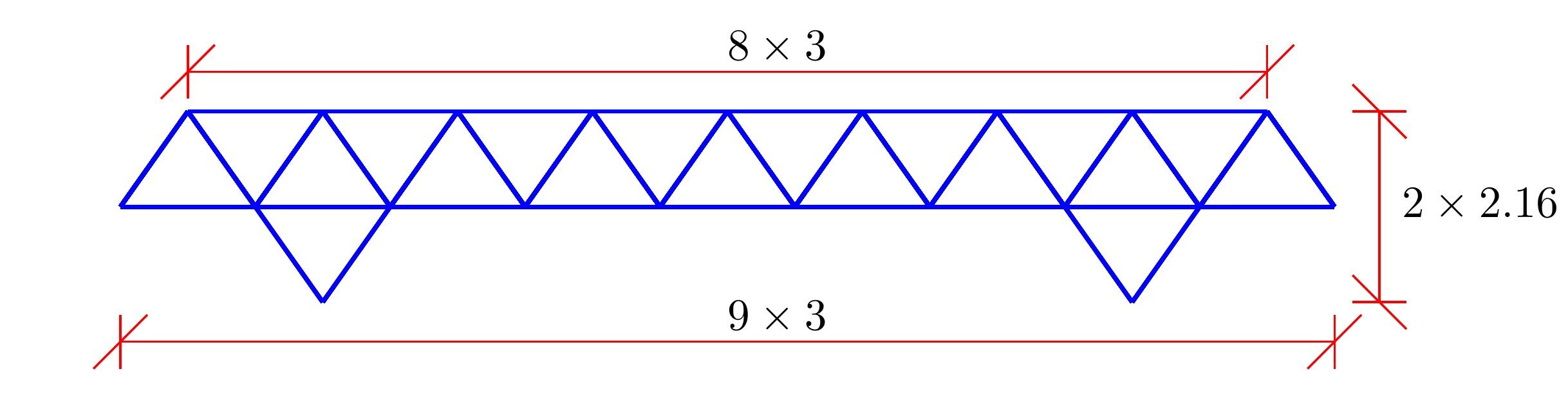}
		\caption{Model of the roof truss.}
		\label{e2_model}
	\end{center}
\end{figure}
In this example, we consider that a stochastic wind load acts vertically downward on a roof truss. As shown in Fig.\ref{e2_model}, the roof truss from \cite{zheng2020anovelsfem} includes 185 spatial nodes and 664 elements, where material properties of all members are set as Young's modulus $E = 209{\mathop{\rm GPa}\nolimits}$ and cross-sectional areas $A = 16{{\mathop{\rm cm}\nolimits} ^2}$. The stochastic wind load $f\left( {x,y,\theta } \right)$ is a random field with the covariance function ${C_{ff}}\left( {{x_1},{y_1};{x_2},{y_2}} \right) = \sigma _f^2{e^{ - {{\left| {{x_1} - {x_2}} \right|} \mathord{\left/ {\vphantom {{\left| {{x_1} - {x_2}} \right|} {{l_x}}}} \right. \kern-\nulldelimiterspace} {{l_x}}} - {{\left| {{y_1} - {y_2}} \right|} \mathord{\left/ {\vphantom {{\left| {{y_1} - {y_2}} \right|} {{l_y}}}} \right. \kern-\nulldelimiterspace} {{l_y}}}}}$, where the variance $\sigma _f^2 = 1.2$, the correlation lengths ${l_x} = {l_y} = 24$, and it can be expanded into a series form by use of Karhunen-Lo\`{e}ve expansion \cite{phoon2002simulation, zheng2017simulation, Rahman2018A} with $M$-term truncated as
\begin{equation} \label{E2_RF}
  f\left( {x,y,\theta } \right) = \sum\limits_{i = 0}^M {{\xi _i}\left( \theta  \right)\sqrt {{\nu _i}} {f_i}\left( {x,y} \right)}
\end{equation}
where $\left\{ {{\xi _i}\left( \theta  \right)} \right\}_{i = 1}^M$ are uncorrelated standard Gaussian random variables, ${{\nu _i}}$ and ${{f_i}\left( {x,y} \right)}$ are eigenvalues and eigenfunctions of the covariance function ${C_{ff}}\left( {{x_1},{y_1};{x_2},{y_2}} \right)$, which can be obtained by solving a eigen equation \cite{zemyan2012classical}, ${\nu _0} = {\xi _0}\left( \theta  \right) \equiv 1$ and the mean function ${f_0}\left( {x,y} \right) = 10{\mathop{\rm kN}\nolimits}$.

Similar to Example \ref{Example1}, we consider the failure probability at the maximum displacement of the roof truss and the limit state function $g\left( \theta  \right)$ is defined by the maximum displacement as
\begin{equation}\label{e2_gx}
  g\left( \theta  \right) = \mathop {\max }\limits_i {u_i}\left( \theta  \right) - c \cdot {u_{mean}}
\end{equation}
where ${u_{mean}} = {\mathop{\rm mean}\nolimits} \left( {\mathop {\max }\limits_i {u_i}\left( \theta  \right)} \right)$ is the mean value of the maximum displacement, ${u_i}\left( \theta  \right)$ are vertical displacements of all spatial nodes, and the scalar $c$ is related to different failure probabilities.

\begin{figure}[!t]
	\begin{center}
		\includegraphics[width=0.7\textwidth]{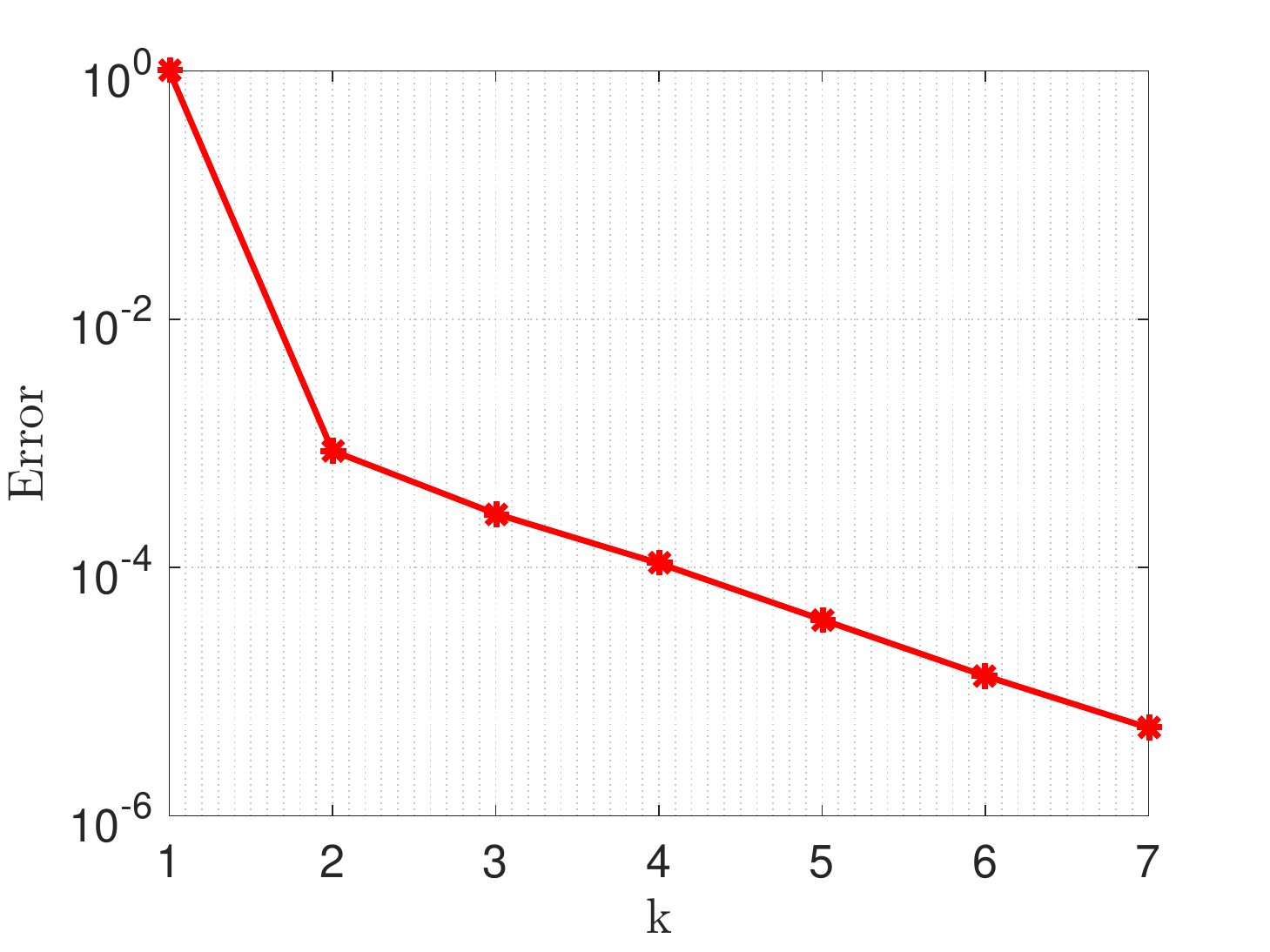}
		\caption{Iteration errors of different retained items.}
		\label{e2_err}
	\end{center}
\end{figure}

\begin{figure}[!t]
	\begin{center}
		\includegraphics[width=1.0\textwidth]{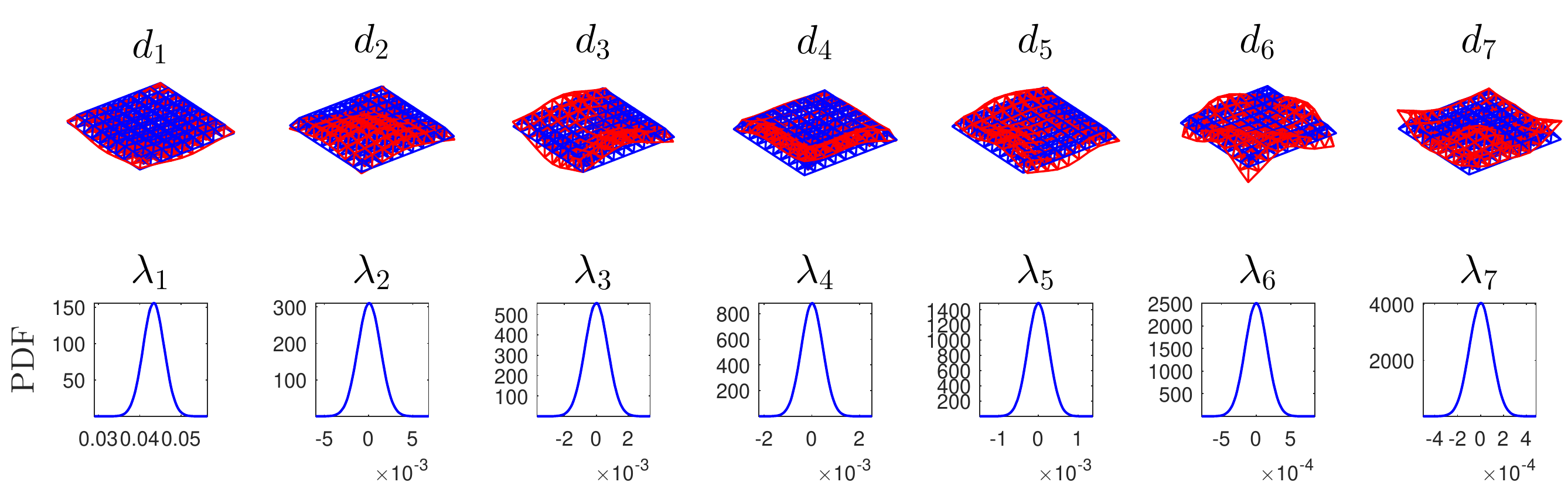}
		\caption{Solutions of the couples $\left\{ {{\lambda _i}\left( \theta  \right),{d_i}} \right\}_{i = 1}^7$.}
		\label{e2_DL}
	\end{center}
\end{figure}

A stochastic finite element equation of the stochastic response $u\left( \theta  \right)$ is obtained based on the expansion Eq.\eqref{E2_RF} of the stochastic wind load. In order to show the effectiveness of the proposed method for high-dimensional reliability analysis, we adopt the stochastic dimension $M=100$ in Eq.\eqref{E2_RF}. It is seen from Fig.\ref{e2_err} that seven iterations can achieve the required precision ${\varepsilon _1} = 1 \times {10^{ - 5}}$, which indicates the fast convergence rate of the proposed method even for very high stochastic dimensions. The deterministic response components $\left\{ {{d_i}} \right\}_{i = 1}^7$ and corresponding random variables $\left\{ {{\lambda _i}\left( \theta  \right)} \right\}_{i = 1}^7$ are shown in Fig.\ref{e2_DL}. The computational time for solving couples $\left\{ {{\lambda _i}\left( \theta  \right),{d_i}} \right\}_{i = 1}^7$ in this example is less than half a minute by use of a personal laptop (dual-core, Intel i7, 2.40GHz), which indicates that Algorithm \ref{alg1} is still less computational costs for high-dimensional stochastic problems.


\begin{figure}[h]
	\begin{center}
		\includegraphics[width=0.7\textwidth]{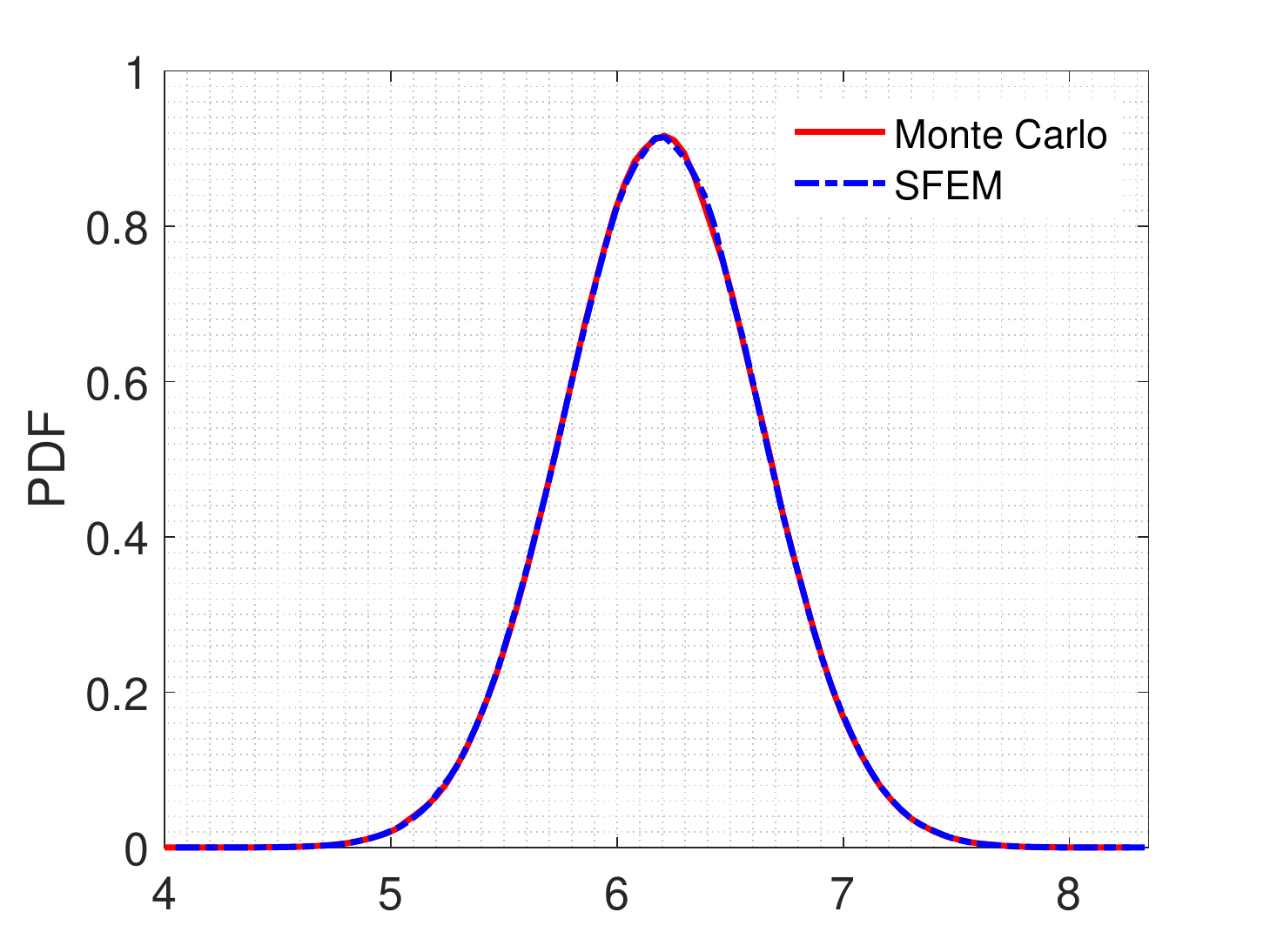}
		\caption{PDF of the maximum displacement}
		\label{e2_pdf}
	\end{center}
\end{figure}

The resulted approximate probability density function (PDF) of the maximum stochastic displacement of the whole roof truss compared with that obtained from the Monte Carlo simulation is seen in Fig.\ref{e2_pdf}, which indicates that the result of seven-term approximation is in very good accordance with that from the Monte Carlo simulation. Further increasing the number of couples won't significantly improve the accuracy since the series in Eq.\eqref{uk} has converged. It is noted that, the tail of the probability distribution is crucial for reliability analysis. The proposed method is sample-based, which provides the possibility for high-precision reliability analysis.

\begin{figure}[h]
	\begin{center}
		\includegraphics[width=0.7\textwidth]{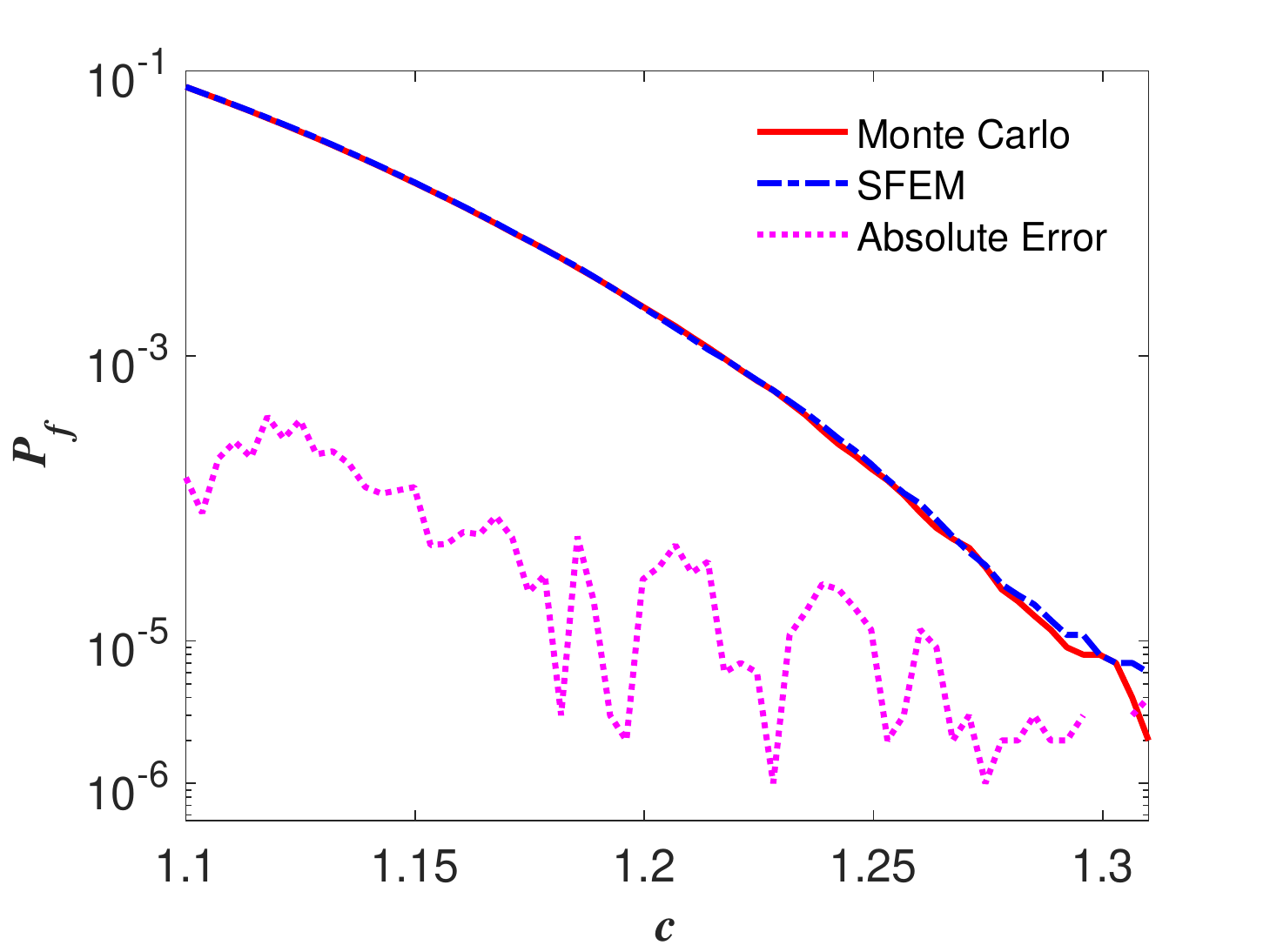}
		\caption{Failure probabilities of different scalar $c$.}
		\label{e2_Pf}
	\end{center}
\end{figure}

In this example, the scalar parameter $c$ in Eq.\ref{e2_gx} is set from 1.10 to 1.31, and failure probabilities of different scalar $c$ are shown in Fig.\ref{e2_Pf}. The accuracy and efficiency of the proposed method is verified again in comparison to the Monte Carlo simulation. The accuracy is reduced when the failure probability $P_f$ is close to $10^{-6}$, but it is still very close to that from the Monte Carlo simulation. In this sense, the Curse of Dimensionality encountered in high-dimensional reliability analysis, can be overcome successfully with cheap computational costs.

\subsection{Global reliability analysis of a plate} \label{Example3}
\begin{figure}[htbp]
	\begin{center}
		\includegraphics[width=0.9\textwidth]{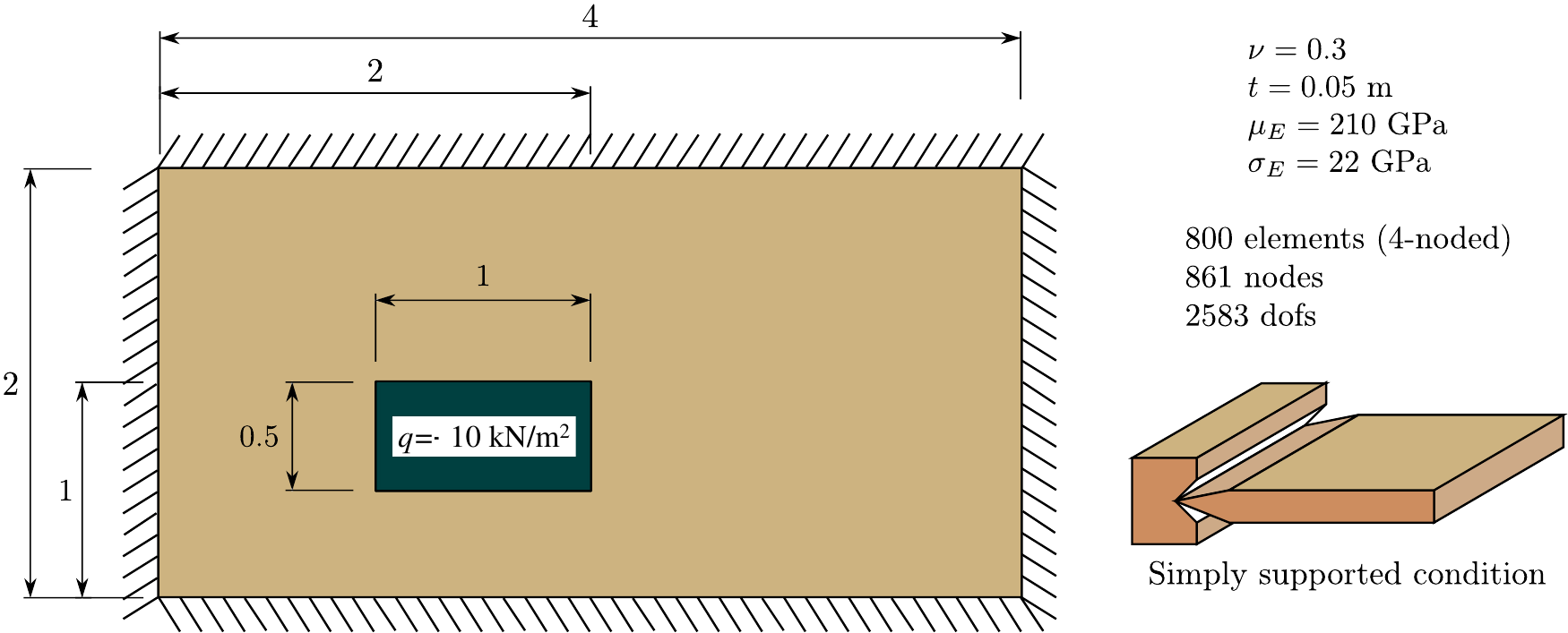}
		\caption{Model of the plate.}
		\label{e3_model}
	\end{center}
\end{figure}

In this example, we consider a Kirchhoff-Love thin plate subjected to a deterministic distributed load $q=-10{{{\rm{kN}}} \mathord{\left/ {\vphantom {{{\rm{kN}}} {{{\rm{m}}^2}}}} \right. \kern-\nulldelimiterspace} {{{\rm{m}}^2}}}$ and simply supported on four edges, which is modified from \cite{uribe2015probabilistic}. As shown in Fig.\ref{e3_model}, parameters of this problem are set as length $L = 4{\rm{m}}$, width $D = 2{\rm{m}}$, thickness $t=0.05{\rm{m}}$ and Poisson's ratio $\nu=0.3$. For the sake of simplicity, we neglect self-weight of the plate and assume Young's modulus $E\left( {x,y,\theta } \right)$ as the realization of a Gaussian random field with mean function ${\mu _E} = 210{\rm{GPa}}$ and covariance function ${C_{EE}}\left( {{x_1},{y_1};{x_2},{y_2}} \right) = \sigma _E^2{e^{ - {{\left| {{x_1} - {x_2}} \right|} \mathord{\left/ {\vphantom {{\left| {{x_1} - {x_2}} \right|} {{l_x}}}} \right.  \kern-\nulldelimiterspace} {{l_x}}} - {{\left| {{y_1} - {y_2}} \right|}  \mathord{\left/ {\vphantom {{\left| {{y_1} - {y_2}} \right|} {{l_y}}}} \right. \kern-\nulldelimiterspace} {{l_y}}}}}$ with correlation lengths ${l_x} = 2{\rm{m}}$, ${l_y} = 4{\rm{m}}$, ${\sigma _E} = 22{\rm{GPa}}$, ${l_x} = 2$m, ${l_y} = 4$m. Similar to Eq.\ref{E2_RF}, Young's modulus $E\left( {x,y,\theta } \right)$ is represented by Karhunen-Lo\`{e}ve expansion with 10-term truncated as
\begin{equation} \label{E3_RF}
  E\left( {x,y,\theta } \right) = {\mu _E} + \sum\limits_{i = 1}^{10}{{\xi _i}\left( \theta  \right){E_i}\left( {x,y} \right)}
\end{equation}

In this example, we consider the failure probabilities of all spatial points, which can be considered as a global reliability analysis. The global limit state function $g\left( {\bf{x}}, \theta  \right)$ is defined by the stochastic displacement of the plate exceeding a critical threshold as, 
\begin{equation}\label{e3_gx}
  g\left( {{\bf{x}},\theta } \right) = {u_\omega }\left( {{\bf{x}},\theta } \right) - c \cdot {u_{\omega,mean}}\left( {\bf{x}} \right)
\end{equation}
where ${u_\omega }\left( {{\bf{x}},\theta } \right)$ is the vertical stochastic displacement field of all spatial nodes, ${u_{\omega,mean}}\left( {\bf{x}} \right) = {\mathop{\rm mean}\nolimits} \left( {{u_\omega }\left( {{\bf{x}},\theta } \right)} \right)$ is corresponding mean displacement field of ${u_\omega }\left( {{\bf{x}},\theta } \right)$, and the scalar parameter is adopted as $c=1.35$.

\begin{figure}[htbp]
	\begin{center}
		\includegraphics[width=0.7\textwidth]{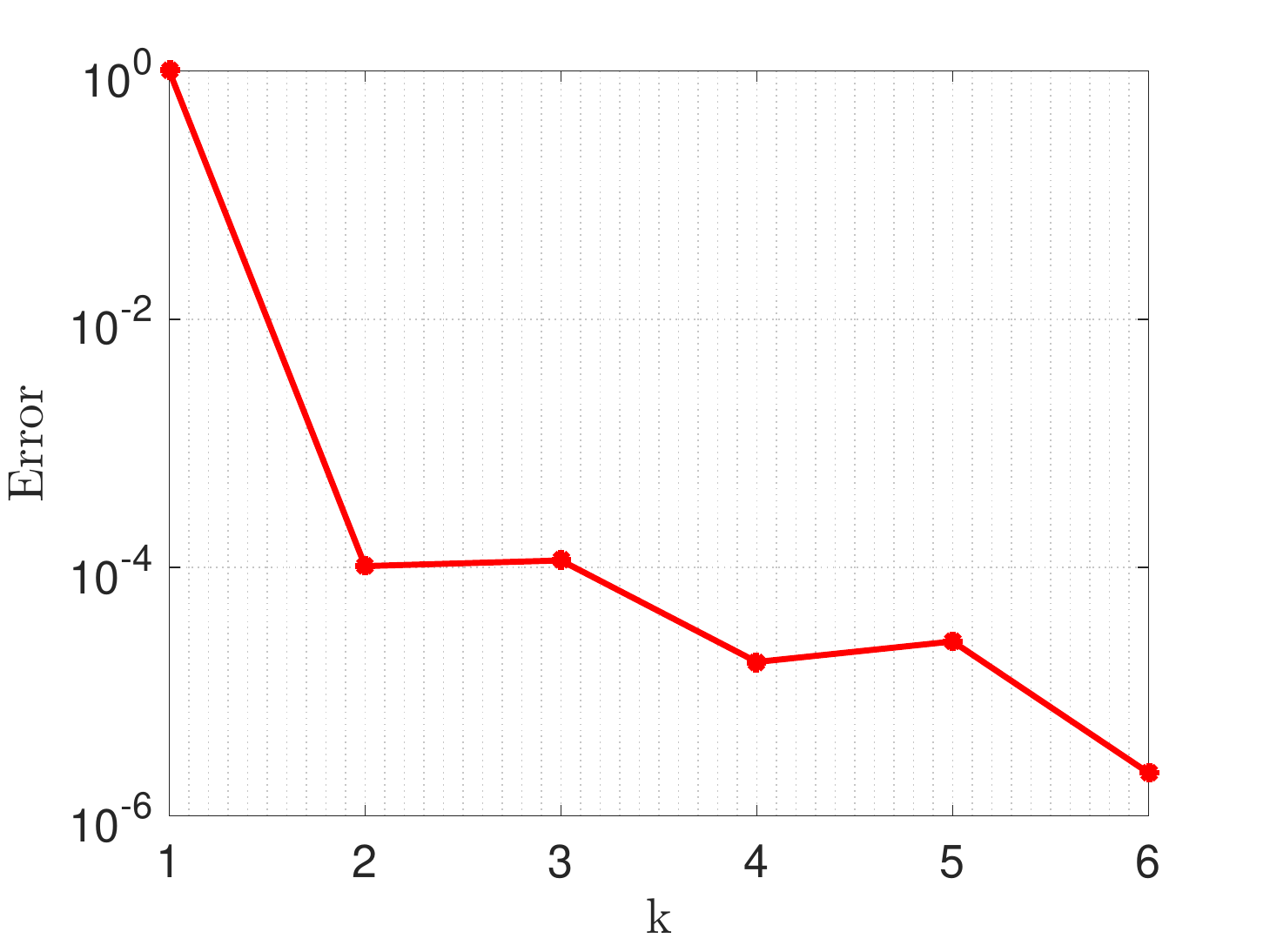}
		\caption{Iteration errors of different retained items.}
		\label{e3_err}
	\end{center}
\end{figure}
\begin{figure}[htbp]
	\begin{center}
		\includegraphics[width=1.0\textwidth]{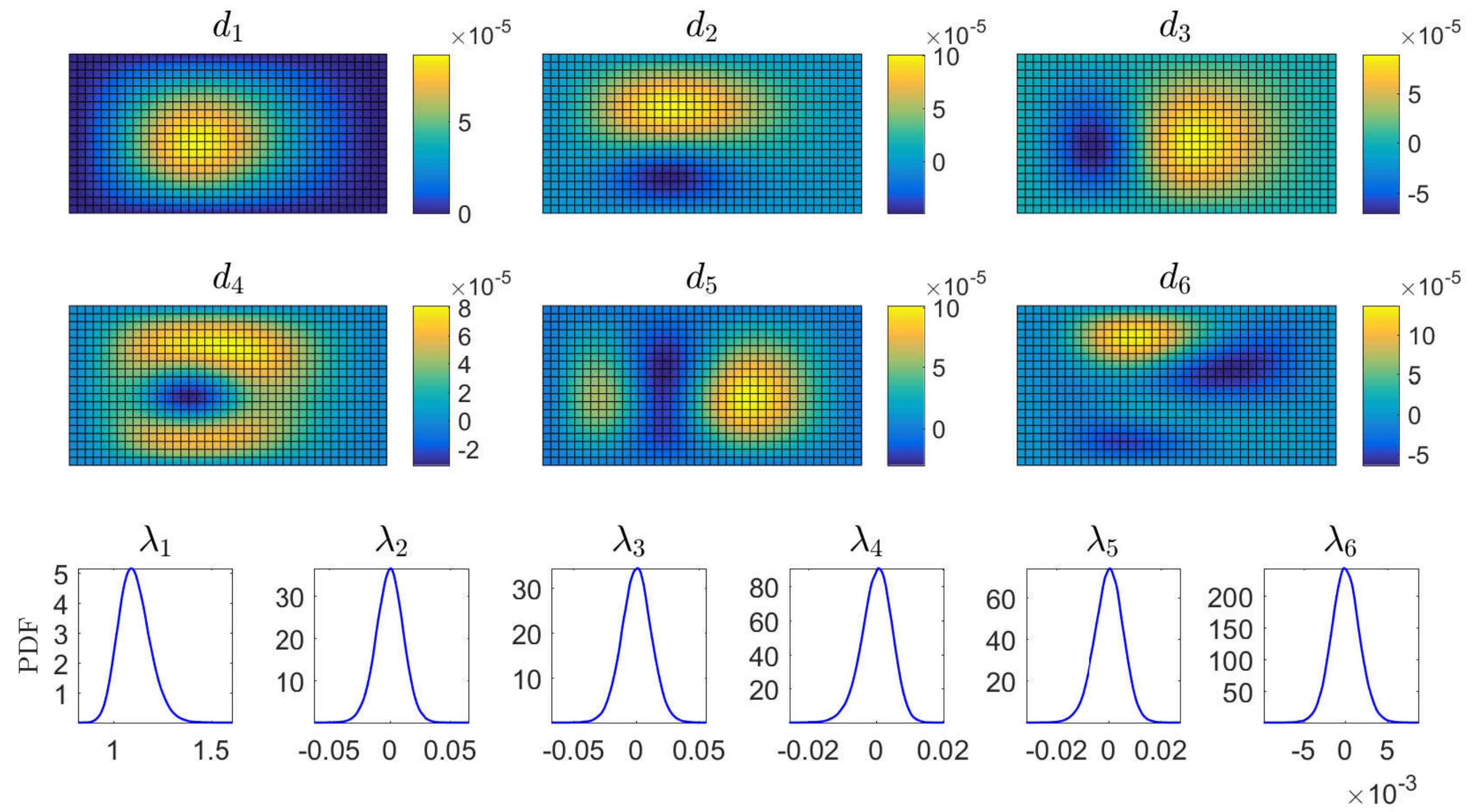}
		\caption{Solutions of the couples $\left\{ {{\lambda _i}\left( \theta  \right),{d_i}} \right\}_{i = 1}^6$.}
		\label{e3_DL}
	\end{center}
\end{figure}

Here we use Kirchhoff-Love finite element theory of plates and four-node Melosh-Zienkiewicz-Cheung (MZC) element to divide the plate into 861 nodes and 800 elements. The unknown node displacement $u(\theta )$ is introduced as $u(\theta ) = {[{u_\omega }(\theta ),{u_x}(\theta ),{u_y}(\theta )]^T}$, which are the vertical displacement, rotations in $x$ and $y$ axes, respectively, then 2583 degrees of freedom are defined and corresponding stochastic finite element equation can be obtained. As shown in Fig.\ref{e3_err}, the required precision ${\varepsilon _1} = 1 \times {10^{ - 5}}$ can be achieved after six iterations, which demonstrate that the proposed method can be applied to large-scale stochastic problems. Fig.\ref{e3_DL} shows the vertical displacement components $\left\{ {{d_i}} \right\}_{i = 1}^6$ and corresponding random variables $\left\{ {{\lambda _i}\left( \theta  \right)} \right\}_{i = 1}^6$, which again indicates that the first few couples dominate the solution even for very complex stochastic problems.

\begin{figure}[htbp]
	\begin{center}
	\includegraphics[width=0.9\textwidth]{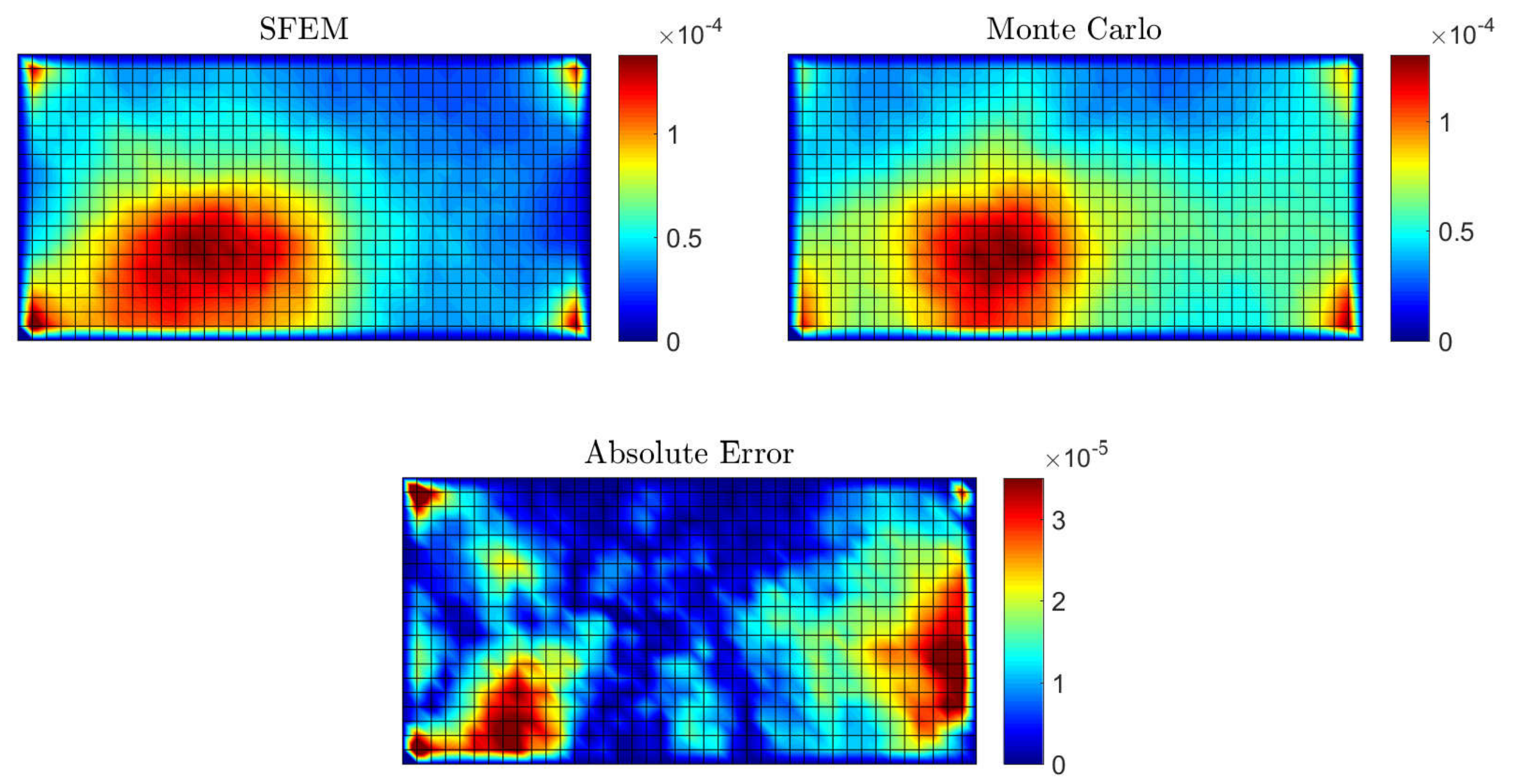}
	\caption{Failure probability nephogram}
	\label{e3_Pf}
	\end{center}
\end{figure}

Based on the vertical stochastic displacement ${u_\omega }\left( {{\bf{x}},\theta } \right)$ obtained by the proposed method, the global failure probability ${P_f}\left( {\bf{x}} \right)$ of the plate can be calculated by use of Eq.\eqref{PfS}, that is, the step \ref{Step11} in Algorithm \ref{alg1}. The failure probability nephogram in a discrete form shown in Fig.\ref{e3_Pf} has a good accordance with that from the Monte Carlo simulation, which demonstrates the effectiveness and accuracy of the proposed method for global reliability analysis. It is noted that, failure probabilities ${P_f}\left( {{x_i}} \right)$ of all spatial nodes constitute the global failure probability ${P_f}\left( {\bf{x}} \right)$ in Fig.\ref{e3_Pf}, thus some difficulties encountered in existing approaches can be circumvented, such as computing the design point (a point lying on the failure surface which has the highest probability density among other points on the failure surface). In this way, the proposed method presents a new strategy for reliability analysis.

\section{Conclusion}\label{Sec6}
This paper proposes an efficient and unified methodology for structural reliability analysis and illustrates its accuracy and efficiency using three practical examples. The proposed method firstly compute structural stochastic responses by using a novel stochastic finite element method and the failure probability is subsequently calculated based on the obtained stochastic responses. As shown in three considered examples, the proposed method has the same implementation procedure for different problems and allows to deal with high-dimensional and large-scale stochastic problems with very low computational costs. The Curse of Dimensionality encountered in high-dimensional reliability analysis can thus be circumvented with great success. In addition, the proposed method gives a high-precision solution of global reliability analysis, which overcomes some difficulties encountered in existing approaches and provides a new strategy for reliability analysis of complex problems. In these senses, the methodology proposed in this paper is particularly appropriate for large-scale and high-dimensional reliability analysis of practical interests and has great potential in the reliability analysis in science and engineering. In the follow-up research, we hopefully further improve the theoretical analysis of the proposed method and apply it to a wider range of problems, such as reliability analysis of time-dependent problems and nonlinear problems \cite{Darmawan2007Spatial, Bichon2012Efficient, Tabbuso2016An}.

\section*{Acknowledgments}
This research was supported by the Research Foundation of Harbin Institute of Technology and the National Natural Science Foundation of China (Project 11972009). These supports are gratefully acknowledged.

\nocite{*}
\bibliography{References}

\end{document}